\documentclass[runningheads]{llncs}
\usepackage{graphicx}
\usepackage{soul}
\usepackage[]{algorithm2e}
\usepackage{xcolor}
\usepackage[caption=false]{subfig}
\begin{document}
\title{Enhancing Block-Wise Transfer with Network Coding in CoAP\thanks{This work has been partially performed in the framework of mF2C project funded by the European Union's H2020 research and innovation programme under grant agreement 730929.}}
%
%
\author{Cao Vien Phung \and
Jasenka Dizdarevic \and
Admela Jukan}
%
\institute{Technische Universit{\"a}t Braunschweig, Germany \\
\email{\{c.phung, j.dizdarevic, a.jukan\}@tu-bs.de}}
\maketitle
\vspace{-0.3cm}
\begin{abstract}
CoAP (Constrained Application Protocol) with block-wise transfer (BWT) option is a known protocol choice for large data transfer in general lossy IoT network environments. Lossy transmission environments on the other hand lead to CoAP resending multiple blocks, which creates overheads. To tackle this problem, we design a BWT with network coding (NC), with the goal to reducing the number of unnecessary retransmissions. The results show the reduction in the number of block retransmissions for different values of blocksize, implying the reduced transfer time. For the maximum blocksize of 1024 bytes and total probability loss of 0.5, CoAP with NC can resend up to 5 times less blocks.

\end{abstract}
\section{Introduction}
One of the most known IoT (Internet of Thing) protocols, CoAP \cite{Shelby}, integrates BWT \cite{Bormann} as a good choice to transmit large amount of data. Since CoAP operates over User Datagram Protocol (UDP) and is thus fundamentally unreliable, it introduces a mode operation \emph{confirmable} where a message is considered delivered once the acknowledgment has been received. This mode is often combined with BWT implementation where a large resources are divided into blocks for transferring. The receiver needs to send an acknowledgment after each received block. In lossy environments, which is typically the case in IoT, these acknowledgments can fail to arrive at the client, resulting in unnecessary retransmissions.

This paper addresses this problem of unnecessary retransmission by combining BWT in CoAP with NC. Similar approach has been done for REST HTTP in \cite{Phung}.
Since REST HTTP and CoAP follow the same request-reply paradigm, the
REST HTTP algorithm was modified for the specific CoAP requirements.
Instead of adding a NC layer for REST, in this paper we introduce a novel design which adds NC technique in form of a so called \emph{option value} for BWT. It is a simple coding scheme with only XOR operations for the normal coded blocks, except the additional blocks using random linear network coding (RLNC) to better operate in constrained devices and environment. The numerical results show that additional retransmission of blocks can be reduced.

\section{Related work}
The authors in \cite{Schutz} extend BWT using NC, while authors in \cite{Choi} propose a scheme where multiple blocks can be retrieved by one request, focusing more on the problem of reducing latency. The goal of these schemes is to reduce communication time. Our paper focuses on another approach, combining NC and BWT based on the work in \cite{Phung} to reduce the amount of traffic that needs to be resent.

\section{Our design}
Our scenario considers a CoAP client server communication as shown in Fig.\ref{scenario}. The client sends a large resource divided into $5$ blocks. Our scheme uses BWT with stop-and-wait mechanism. BWT without NC in Fig.\ref{scenario}.1 allows the blocks to be retransmitted when the client does not receive their acknowledgment in timeout interval. However, resending blocks $p_1$ is unnecessary because it has arrived at the server. To address this issue, we design a NC scheme in Fig.\ref{scenario}.2. We observe the acknowledgment of block $p_1$ is lost, but the client is unaware of what is happening at the server. So, the client should perform NC among blocks after each timeout. In our scenario, one new block is only presented by one coded block at a time. Therefore, coded blocks are always linearly independent \cite{Van}. Performing NC with only XOR operations is enough. With simple XOR operations, we can remove coding coeffcients from the option value. As a result,we can dramatically reduce the protocol overhead. At the time of arriving coded block $(p_1+p_2+p_3+p_4)$, along with block $p_1$ received before, the server operates Gauss Jordan Elimination(GJE) to identify seen blocks $p_1$ and $p_2$ (refer to \cite{Phung} to understand seen packets). The acknowledgment R($sn$,$htp$,$rdt_s$)=(2,4,2) can be responded even when the original blocks have not yet been decoded, where $sn=2$, $htp=4$ and $rdt_s=2$ are the newest seen block, highest block $ID$ that the server has, and   number of additional blocks, respectively. The two additional blocks ($\delta_1 p_3+\delta_2 p_4$) and ($\delta_3 p_3+\delta_4 p_4$) are resent using RLNC, since they are coded from the previous blocks. The first additional block is lost. When R($3$,$4$,$1$) comes, based on  the option value, the client can identify this one responded from the second additional block, and decide to send the native block $p_4$ instead of coded block $\delta_5 p_4+\delta_6 p_5$ as \cite{Phung} to decrease coding/decoding complexity. Observe that BWT with NC can shorten $1$ block cycles compared to BWT.

\subsection{Option value}
\begin{figure*}[ht]
  \centering
  \subfloat[Scenario of BWT with and without NC]{\includegraphics[ width=6cm, height=3cm]{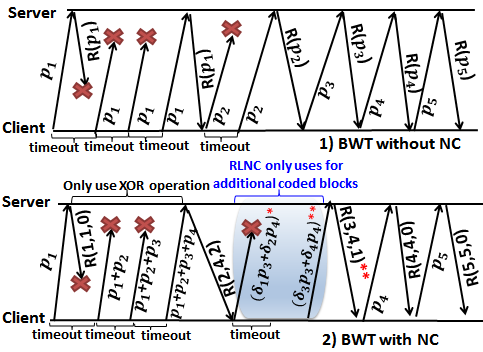}
  \label{scenario}}
  \subfloat[Option value]{\includegraphics[ width=5cm, height=2.2cm]{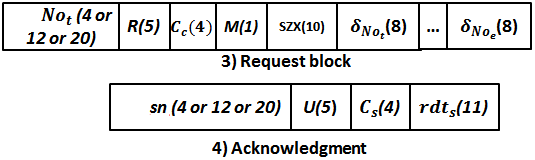}
  \label{option}}
  \caption{Scenario BWT with and without NC, and option value for BWT with NC.}
  \label{additionalblocks}
  \vspace{-0.2cm}
  \end{figure*}
Fig.\ref{option} shows the option value of request block. The typical sizes (in bits) of various fields are written inside. $No_t$ and $No_e$ are the minimum and maximum block index, respectively, involved in the random linear combination (RLC), where $No_e=R+No_t$. $C_c$ is the number tagged for each coded block to distinguish the acknowledgment of which the transferred block is. $M=0$ and $M=1$ (more flag) show the coded block contains and does not contain the last block, respectively. $SZX$ is the block size. $\delta_{i}$ is the coefficient of $i^{th}$ block. Fig.\ref{option} also shows the option value of acknowledgment. $C_s$ is copied from $C_c$. $rdt_s$, $htp$ and $sn$ were defined above, where $htp$ is indirectly represented via $U$, where $htp=U+sn$.
\subsection{Coding and decoding, and computing additional blocks}
The coding is similarly performed as \cite{Phung}, but one new feature of Algorithm \ref{al} is added to distinguish which acknowledgement responds for corresponding block.

\begin{algorithm}[H]
\SetAlgoLined

  \eIf{Acknowledgement (Ack) received for the additional blocks}{
      \eIf{$C_s<C_c$}{
    Ack of previous additional block; perform as \cite{Phung}, but use XOR;
    }
    {
    \textbf{if} $rdt_s>0$, detect losses; Resend using RLNC as \cite{Phung};
    }
   }{
   \textbf{if} $htp<No_e$, Ack of previous normal block: No transmission;
  }

 \caption{Acknowledgement identification.}
 \label{al}
\end{algorithm}

For decoding, acknowledgement method, decoding and delivery method, buffer management method are similarly performed same as \cite{Phung}.

Let $N$, $R$ and $B$ be the total number of blocks of a resource, size of resource and of each block, respectively. The total number of blocks sent is $N=\left \lceil R/B \right \rceil$. Based on analysis in \cite{Phung}, the number of additional blocks of BWT $A_{WoNC} = (N/(1-p)) - N$  and of NC\_BWT $A_{WNC} = (N/(1-(\alpha\cdot p))) - N$, where $p$ is the total loss probability for both request block and acknowledgement, and $\alpha$ is the loss rate when the client transfers block to the server.

\section{Numerical results}
\begin{figure*}[ht]
    \vspace{-0.7cm}
  \centering
  \subfloat[$\alpha=0.3$]{\includegraphics[ width=4cm, height=3.5cm]{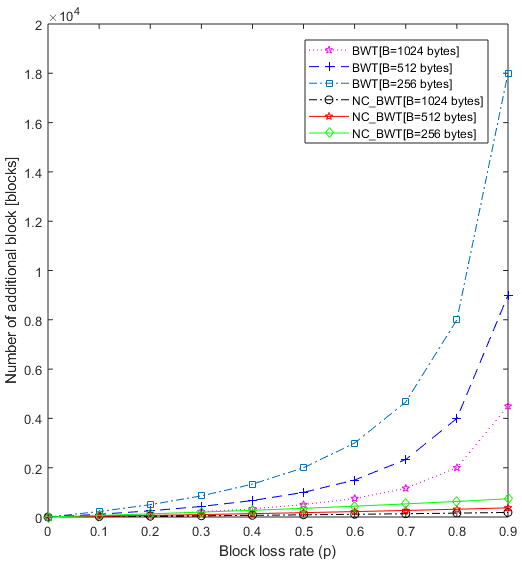}
  \label{1}}
  \subfloat[$\alpha=0.7$]{\includegraphics[ width=4cm, height=3.5cm]{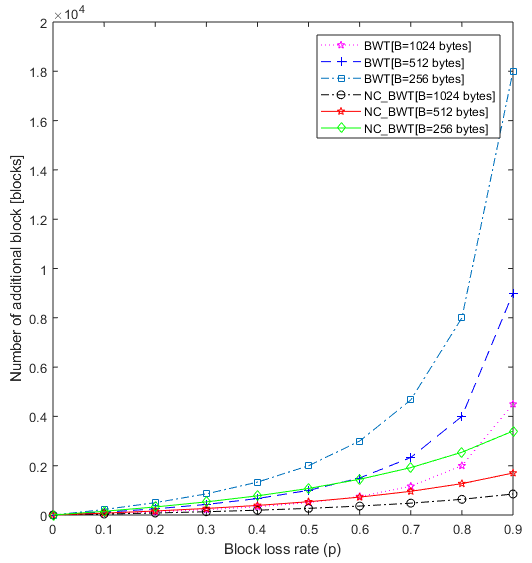}
  \label{2}}
  \subfloat[$\alpha=1$]{\includegraphics[ width=4cm, height=3.5cm]{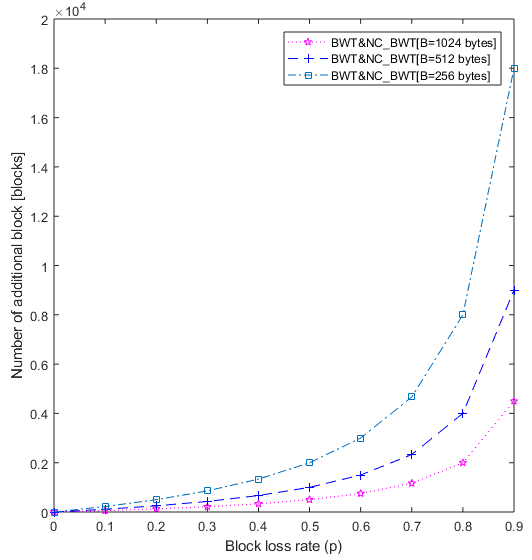}
  \label{3}}
  \caption{Number of additional blocks with network coding NC\_BWT and without BWT}
  \label{additionalblocks}
  \vspace{-0.7cm}
  \end{figure*}

This section shows numerical results to compare NC\_BWT with BWT in term of the number of additional blocks in Fig.\ref{additionalblocks}. We consider an application with $R=512KB$, where $3$ types of block size $B$ are chosen: $1024$ bytes, $512$ bytes, and $256$ bytes. The loss probability $p$ is considered in $[0;0.9]$. Three values of the request block loss rate $\alpha=0.3$; $0.7$ and $1$ are selected. $A_{WoNC} = (N/(1-p)) - N$ and $A_{WNC} = (N/(1-(\alpha\cdot p))) - N$ are used to compute the number of additional blocks for BWT and NC\_BWT, respectively. We see that for both BWT and NC\_BWT when loss probability $p$ increases, the number of additional blocks also increases. We also observe that the number of retransmissions with block size $B=256$ bytes is the highest for both BWT and NC\_BWT because under the impact of block loss, if a smaller block size is selected, the resource is divided into more blocks, therefore leading to more block losses, and resulting in more retransmissions. NC\_BWT always outperforms BWT term of the number of additional blocks. We consider an example of $p=0.5$, $B=1024$ bytes, BWT needs to resend $500$ blocks for all values $\alpha$, but NC\_BWT only resends $88.235$ blocks for $\alpha=0.3$,  and $269.231$ blocks for $\alpha=0.7$. In addition, we observe that the smaller the loss rate value $\alpha$, the more the benefit from NC\_BWT is. Fig.\ref{3} ($\alpha=1$) shows that, NC\_BWT does not have any benefit from NC for all $p$.

\section{Conclusion}
In this paper, we consider a combination between BWT with NC in CoAP. We shows how our algorithm can reduce the number of additional blocks. In future works, we will do simulation to see the impact of NC on large resource transfer.

\end{document}